# The Credibility Theory applied to backtesting Counterparty Credit Risk

**Matteo Formenti**

*Group Risk Management*
*UniCredit Group*

*Università Carlo Cattaneo*

September 3, 2014

**Abstract**

Credibility theory provides tools to obtain better estimates by combining individual data with sample information. We apply the Credibility theory to a Uniform distribution that is used in testing the reliability of forecasting an interest rate for long term horizons. Such empirical exercise is asked by Regulators (CRR, 2013) in validating an Internal Model Method for Counterparty Credit Risk. The main results is that risk managers consider more reliable the output of a test with limited sample size when the Credibility is applied to define a confidence interval.

## Introduction: A short review of Credibility Theory

The classical Credibility Theory was first introduced by Mowbray (1914) in the definition of the equal premium in the insurance market. The credibility is defined as the weight to apply to the individual observation when calculating the estimate of a variable. The basic formula for calculating credibility weighted estimate is:

$$Estimate = Z\,[Individual\ Observation] + (1-Z)[Other\ Information]$$

where $Z \in (0,1)$ is the credibility weight assigned to the individual observation, and $(1-Z)$ is generally referred to as the complement of credibility assigned to the observations extracted from the sample the individual observation is belonging to. The key is, thus, to determine $Z$ that is the reliability of the individual observation and to produce a better estimate with the help of credibility[1].

Note that if the body of observed data is large and not likely to vary much from one period to another, then $Z$ will be closer to one. On the other hand, if the sample consists of limited data, then $Z$ will be closer to zero and more weight will be given to other information. As a consequence, Credibility Theory can be a reliable tool to assess the estimate computed with limited sample size and how much credibility should be assigned to the information known.

The credibility $Z$ can been determined through (i) the Classical Credibility, where $Z$ is a function of the expected variance of the observations versus the individual variance, or (ii) the Bayesian analysis developed by Bühlmann that combines the current observations with prior information[2].

In Classical Credibility the goal is to obtain how much data one needs before to assign to the individual observation the 100% credibility. This amount of data is usually referred to as the *Full Credibility Criterion* or the *Standard for Full Credibility*[3]. In order to determine the Full Credibility for a frequency, that is usually modeled by a Poisson process, the Normal Approximation can be used to estimate **how often the observed results will be far from the mean**.

The probability $P$ that the observation $X$ is within $\pm k$ from the population mean $\mu$ is:

$$P = Prob[\mu - k\mu \leq X \leq \mu + k\mu]$$
$$= Prob\left[-k\left(\frac{\mu}{\sigma}\right) \leq \frac{X-\mu}{\sigma} \leq +k\left(\frac{\mu}{\sigma}\right)\right]$$

where the last expression assuming the Normal Approximation let us use the table of the Standard Normal Distribution. It is important to remark that the $(\mu, \sigma)$ are the mean and standard deviation of the frequency process. In the Poisson case where the frequency is $n$, we have $\mu = n$; $\sigma = \sqrt{n}$ and the probability that **the observed frequency $X$ is within $\pm k$ of the expected number $\mu = n$** equals:

$$P = Prob\left[-k\left(\frac{\mu}{\sigma}\right) \leq \frac{X-\mu}{\sigma} \leq +k\left(\frac{\mu}{\sigma}\right)\right]$$
$$= Prob\left[-k\sqrt{n} \leq \frac{X-\mu}{\sigma} \leq +k\sqrt{n}\right]$$

In terms of the cumulative distribution for the unit $\Phi(u)$ where $u = \frac{X-\mu}{\sigma}$ we have:

---

[1] Assume a sample of drivers and the frequency of their accidents. If the insurance company want to customize the insurance premium could determine the individual premium by weighting (Z) the individual frequency of accident with the average frequency of accident by (1-Z).
[2] The Bühlmann credibility estimates are also called the best linear least squares fits to Bayesian credibility
[3] If the amount of data is the Full Credibility Criterion then $Z = 1$, otherwise $Z \in (0,1)$).

$$P = \Phi(k\sqrt{n}) - \Phi(-k\sqrt{n}) = \Phi(k\sqrt{n}) - (1 - \Phi(k\sqrt{n}))$$
$$= 2\Phi(k\sqrt{n}) - 1$$

or equivalently $\Phi(k\sqrt{n}) = \frac{1+P}{2}$. Table 1 shows the probability of being within $\pm k$ of the mean assuming the Standard Normal distribution.

| # frequency | k=10% | k=5% | k=2.5% | k=1% | k=0.5% |
|---|---|---|---|---|---|
| 10 | 24,82% | 12,56% | 6,30% | 2,52% | 1,26% |
| 50 | 52,05% | 27,63% | 14,03% | 5,64% | 2,82% |
| 100 | 68,27% | 38,29% | 19,74% | 7,97% | 3,99% |
| 500 | 97,47% | 73,64% | 42,38% | 17,69% | 8,90% |
| 1000 | 99,84% | 88,62% | 57,08% | 24,82% | 12,56% |
| 5000 | 100,00% | 99,96% | 92,29% | 52,05% | 27,63% |
| 10000 | 100,00% | 100,00% | 98,76% | 68,27% | 38,29% |

Table 1 Probability of being within ±k of the mean (Normal Distribution)

We can exploit the equation above by changing the objective of our analysis and considering as **exogenous** the probability $P$. In fact, it is possible to **compute the number of expected frequency $(n_0)$ such that the change of being within $\pm k$ of the mean is P**. In fact $(n_0)$ can be calculated from the previous formula calling for simplicity:

$$y = k\sqrt{n}$$

that is determined from a normal table once we fixed **P**. Solving for $(n_0)$ we get:

$$\Phi(k\sqrt{n_0}) = \frac{1+P}{2}$$
$$n_0 = \frac{y^2}{k^2}$$

where y is such that $\Phi(y) = \frac{1+P}{2}$, and $n_0$ is the Full Credibility Criterion. Table 2 shows the Full Credibility for the frequency $n_0$ given different values of P and k.

| # sample | k=30% | k=20% | k=10% | k=5% | k=1% |
|---|---|---|---|---|---|
| 80,00% | 18 | 41 | 164 | 657 | 16.424 |
| 90,00% | 30 | 68 | 271 | 1.082 | 27.055 |
| 95,00% | 43 | 96 | 384 | **1.537** | 38.415 |
| 97,50% | 56 | 126 | 502 | 2.010 | 50.239 |
| 99,00% | 74 | 166 | 663 | 2.654 | 66.349 |
| 99,50% | 88 | 197 | 788 | 3.152 | 78.794 |
| 99,99% | 168 | 378 | 1.514 | 6.055 | 151.367 |

Table 2 Full Credibility Criterion for Normal Distribution as approximation of Poisson Distribution

Now if we want to know how many observations are needed to be "sure" at 95% ($P = 95\%$) of having an observation (a frequency in our example) that is within $\pm 5\%$ from the real mean is $P = 95\%, y = 1.960, \Phi(1.960) = 97.5\%$, $n_0 = \frac{y^2}{k^2} = \frac{1.96^2}{0.05^2} = 1537$.

The Full Credibility Criterion is a powerful tools but with some limits, as it depends on the assumptions: (i) that the observations are frequency, (ii) the frequency is driven by a Poisson process, (iii) there are enough observations to use the Normal Approximation to the Poisson process. In order to have a more general approach, Mayerson (1968) derived a general formula when the Poisson distribution does not apply. The Full Credibility Criteria for frequency becomes:

$$\Phi(k\sqrt{n_0}) = \frac{1+P}{2}$$
$$n_0 = \frac{y^2}{k^2}\left(\frac{\tilde{\sigma}^2}{\tilde{\mu}}\right)$$

where $(\tilde{\mu}; \tilde{\sigma})$ are corresponding two moment of the distribution assumed for the sample[4]. The Mayerson equation has been also exploit to estimate the average size of frequency. Assume a sample of $N$ frequency independently drawn from a distribution with mean $\mu_s$ and variance $\sigma_s^2$. The sample mean of the distribution can be estimated as $(X_1 + X_2 + \cdots + X_N)/N$, and the sample variance as $VAR((\sum X_i)/N) = (\sigma_s^2)/N$. Therefore the probability that the observed sample mean is within $\pm k$ of the distribution mean $\mu_s$ is:

$$P = Prob[\mu_s - k\mu_s \leq S \leq \mu_s + k\mu_s]$$
$$= Prob\left[-k\sqrt{N}\left(\frac{\mu_s}{\sigma_s}\right) \leq \frac{X-\mu}{\sigma} \leq +k\sqrt{N}\left(\frac{\mu_s}{\sigma_s}\right)\right]$$

where the equation is identical to the previous one, with the exception of the additional factor $(\mu_s/\sigma_s)$. According to the Central Limit Theorem the sample distribution can be approximated by a normal distribution for large $N$. So assuming the Normal distribution in order to have a probability P that the observed sample mean will differ from the real mean by less than $\pm k\mu_s$ we have

$$y = k\sqrt{N}\left(\frac{\mu_s}{\sigma_s}\right)$$

where solving for $N$ we obtain the Full Credibility Criterion for frequency computed from any Distribution

$$N = \left(\frac{y}{k}\right)^2 \left(\frac{\sigma_s}{\mu_s}\right)^2$$
$$= n_0 CV_s^2$$

where CV, the ratio of the standard deviation to the mean, is the coefficient of variation of the frequency size distribution, and $n_0$ is the Full Credibility Criterion for a given Probability $P$ and range $k$.

### Counterparty Credit Risk: the Backtesting empirical case

Backtesting refers to the quantitative comparison of a model's forecasts against realized values with the goal of detecting potential weakness of internal models estimation of future quantity, such as interest rates, commodity or mark-to-market of OTC derivatives, used for risk management purposes.

CRR (2013) states that "the institution shall carry out backtesting using historical data on movements in market risk factors […]. That backtesting shall consider a number of **distinct prediction time horizons out to**

---
[4] Note that The extra-factor is reduced in the Poisson case to one.

**at least one year**, over a range of various initialization dates and covering a wide range of market conditions" (art. 294 1(a)). Moreover, as remarked by Basel Committee (2010) "the significance of the results depends for the most part on the amount and quality of data used". This is why Regulators encourage bank to collect an amount of dataset using **non-overlapping** forecast **intervals**. Finally remark that Regulators let the banks choose the appropriate method to aggregate and then validate the overall quality of the forecasts.

The main consequence of using non-overlapping dataset is that any test, used to judge the quality of forecasting time horizons higher than one year, will have less than ten observations for ten years of non-overlapping dates. The backtesting dataset for long time horizons is intrinsic limited in the sample size because the calibration of the model parameters is of at least three years (as required by CRR art. 292 (2)), and several risk factors, such as interest rates, forex and currency, have limited length of available market data histories (i.e. forex USDEUR started in January, 4 1999).

In this work we applied the Credibility Theory to the results of test used in the backtesting exercise, in order to assess quantitatively the reliability of such results when the sample is limited. We focus on the uniformity test, such as the Kolmogorov-Smirnov (1948), Jarque-Brera or Cramer-von Mises (1962) or Anderson-Darling (1954), because they are the market practice in the banks. This is because, at every simulation date, the realized value is compared to the forecasted distribution and then transformed into quantile. And the time series of the quantiles is tested with a uniform distribution by uniformity test.

The main issue is the reliability of an uniform test with five to ten observations. A priori the significance of the results is very low but, as requested by Regulators, the reliability must be considered. Table 3 shows the pvalues of the Anderson-Darling uniform test of the EUR interest rates curve modeled with a standard interest rate model (i.e. CIR or Black-Karasinski) in the period January 2002- June 2013.

| Risk Factor | Pvalues for Time Horizons | | | | | | | Sample of Time Horizons | | | | | | |
|---|---|---|---|---|---|---|---|---|---|---|---|---|---|---|
| Interest rates | 2w | 1m | 3m | 6m | 1y | 18m | 2y | 2w | 1m | 3m | 6m | 1y | 18m | 2y |
| ZEROEUR | 0,2% | 3,3% | 2,0% | 9,9% | 4,6% | 1,6% | 2,4% | 137 | 136 | 45 | 22 | 11 | 6 | 5 |
| ZEROEUR-06M | 11,0% | 0,5% | 9,4% | 2,4% | 9,9% | 0,4% | 0,8% | 137 | 136 | 45 | 22 | 11 | 6 | 5 |
| ZEROEUR-12M | 3,4% | 0,3% | 2,7% | 1,7% | 6,4% | 0,9% | 1,2% | 137 | 136 | 45 | 22 | 11 | 6 | 5 |
| ZEROEUR-24M | 6,9% | 0,5% | 2,0% | 1,5% | 9,7% | 9,5% | 12,5% | 137 | 136 | 45 | 22 | 11 | 6 | 5 |
| ZEROEUR-03Y | 4,4% | 0,4% | 1,7% | 1,4% | 8,7% | 11,8% | 11,2% | 137 | 136 | 45 | 22 | 11 | 6 | 5 |
| ZEROEUR-05Y | 4,3% | 0,3% | 1,3% | 2,5% | 9,0% | 12,4% | 12,2% | 137 | 136 | 45 | 22 | 11 | 6 | 5 |
| ZEROEUR-10Y | 4,2% | 0,7% | 1,3% | 3,1% | 8,4% | 7,1% | 8,5% | 137 | 136 | 45 | 22 | 11 | 6 | 5 |
| ZEROEUR-15Y | 6,0% | 2,8% | 4,4% | 11,0% | 16,4% | 9,3% | 10,4% | 137 | 136 | 45 | 22 | 11 | 6 | 5 |
| ZEROEUR-20Y | 10,4% | 4,4% | 7,3% | 16,2% | 27,6% | 11,0% | 14,8% | 137 | 136 | 45 | 22 | 11 | 6 | 5 |
| ZEROEUR-30Y | 5,5% | 9,3% | 12,9% | 10,6% | 37,0% | 7,6% | 18,8% | 137 | 136 | 45 | 22 | 11 | 6 | 5 |
| ZEROEUR-50Y | 12,3% | 7,3% | 23,8% | 2,4% | 35,8% | 4,9% | 28,4% | 137 | 136 | 45 | 22 | 11 | 6 | 5 |

**Table 3 Interest rate EUR Curve January 2002- June 2013**

The results of long term horizon shows a very limited sample, as the one year time horizon has eleven observations and the two years has just five observations. Therefore the main issue regards the reliability of the pvalues of those long term horizons **considering the quality of the model forecast will be assessed through them.**

The Credibility Theory can be applied to the results of backtesting. The following equations show the Full Credibility Criterion assuming a Uniform continuous distribution instead of a Normal Approximation (see Table 1). A continuous uniform distribution on the interval (0,1) has the following two moments:

$$\mu_S = \frac{a+b}{2} = \frac{1}{2}, \qquad \sigma_S^2 = \frac{(b-a)^2}{12} = \frac{1}{12}$$

where a=1, b=0. The application to previous equations leads to

$$y = k\sqrt{N}\left(\frac{\mu_S}{\sigma_S}\right) = k\sqrt{N}\left(\frac{\sqrt{12}}{2}\right) = k\sqrt{3N}$$

whereby we get the probability that the observed statistics is within $\pm k$ from the distribution mean $\mu_s$

$$P = \Phi(k\sqrt{3N}) - \Phi(-k\sqrt{3N}) = \Phi(k\sqrt{3N}) - (1 - \Phi(k\sqrt{3N}))$$
$$= 2\Phi(k\sqrt{3N}) - 1$$

| # frequency | k=10% | k=5% | k=2.5% | k=1% | k=0.5% |
|---|---|---|---|---|---|
| 10 | 41,61% | 21,58% | 10,89% | 4,37% | 2,18% |
| 50 | 77,93% | 45,97% | 24,05% | 9,75% | 4,88% |
| 100 | 91,67% | 61,35% | 33,50% | 13,75% | 6,90% |
| 500 | 99,99% | 94,72% | 66,71% | 30,15% | 15,35% |
| 1000 | 100,00% | 99,38% | 82,91% | 41,61% | 21,58% |
| 5000 | 100,00% | 100,00% | 99,78% | 77,93% | 45,97% |
| 10000 | 100,00% | 100,00% | 100,00% | 91,67% | 61,35% |

Table 4 Probability of being within ±k of the mean (Uniform Distribution)

It is then possible to solve for $N$ to obtain the Full Credibility Criterion for frequency extracted from the Uniform Distribution:

$$N = \left(\frac{y}{k}\right)^2 \left(\frac{\sigma_S}{\mu_S}\right)^2$$
$$= n_0 \frac{4}{12} = \frac{n_0}{3}$$

where $N$ is the Full Credibility that depends on $n_0$. Table 5 reports the $N$ for different values of P and k. For example, we can infer that any estimation of mean computed from a sample of N=6 data has a 80% Probability of being in a range of ±30% from the real mean, while for $P = 90\%$ and $k = 10\%$ the sample having 100% Credibility must be of 90 observations.

| # sample | k =30 % | k =20 % | k =10 % | k =5 % | k =1 % |
|---|---|---|---|---|---|
| 80,00% | 6 | 14 | 55 | 219 | 5.475 |
| 85,00% | 8 | 17 | 69 | 276 | 6.908 |
| 90,00% | 10 | 23 | 90 | 361 | 9.018 |
| 95,00% | 14 | 32 | 128 | 512 | 12.805 |
| 97,50% | 19 | 42 | 167 | 670 | 16.746 |
| 99,00% | 25 | 55 | 221 | 885 | 22.116 |
| 99,99% | 56 | 126 | 505 | 2.018 | 50.456 |

Table 5 Full Credibility Criterion for Uniform Distribution

The Full Credibility Criterion let us define a quantitative Credibility weight as the ratio of the real sample size and the Full Credibility Criterion:

$$Z = \frac{n^{Backtesting\ sample}}{N^{Full\ Credibility\ Criterion}}$$

Table 6 shows the credibility for the 6-Month, 1-Year, 18-Months and 2-Year time horizons where are displayed in red the weights higher than 30%. The interpretation of the weights higher than 100% equals the ones with 100%. A common choice is $P = 90\%$ and $k = 10\%$. As a consequence, the reliability of the pvalues of time horizons 6-Month is 24%, the ones of 1-Year is 12% and the 18-Months and 2-Years are 2% and 1%.

| Time Horizon 6-Months (n=22) | | | | | | Time Horizon 1-Year (n=11) | | | | | |
|---|---|---|---|---|---|---|---|---|---|---|---|
| **Prob/k** | **k=30%** | **k=20%** | **k=10%** | **k=5%** | **k=1%** | **Prob/k** | **k=30%** | **k=20%** | **k=10%** | **k=5%** | **k=1%** |
| **80%** | 367% | 157% | 40% | 10% | 0% | **80%** | 183% | 79% | 20% | 5% | 0% |
| **85%** | 275% | 129% | 32% | 8% | 0% | **85%** | 138% | 65% | 16% | 4% | 0% |
| **90%** | 220% | 96% | 24% | 6% | 0% | **90%** | 110% | 48% | 12% | 3% | 0% |
| **95%** | 157% | 69% | 17% | 4% | 0% | **95%** | 79% | 34% | 9% | 2% | 0% |
| **97,5%** | 116% | 52% | 13% | 3% | 0% | **97,5%** | 58% | 26% | 7% | 2% | 0% |
| **99%** | 88% | 40% | 10% | 2% | 0% | **99%** | 44% | 20% | 5% | 1% | 0% |
| **99,99%** | 39% | 17% | 4% | 1% | 0% | **99,99%** | 20% | 9% | 2% | 1% | 0% |

| Time Horizon 18-Months (n=6) | | | | | | Time Horizon 2-Years (n=5) | | | | | |
|---|---|---|---|---|---|---|---|---|---|---|---|
| **Prob/k** | **k=50%** | **k=40%** | **k=30%** | **k=10%** | **k=5%** | **Prob/k** | **k=50%** | **k=40%** | **k=30%** | **k=10%** | **k=5%** |
| **80%** | 100% | 43% | 11% | 3% | 0% | **80%** | 83% | 36% | 9% | 2% | 0% |
| **85%** | 75% | 35% | 9% | 2% | 0% | **85%** | 63% | 29% | 7% | 2% | 0% |
| **90%** | 60% | 26% | 7% | 2% | 0% | **90%** | 50% | 22% | 6% | 1% | 0% |
| **95%** | 43% | 19% | 5% | 1% | 0% | **95%** | 36% | 16% | 4% | 1% | 0% |
| **97,5%** | 32% | 14% | 4% | 1% | 0% | **97,5%** | 26% | 12% | 3% | 1% | 0% |
| **99%** | 24% | 11% | 3% | 1% | 0% | **99%** | 20% | 9% | 2% | 1% | 0% |
| **99,99%** | 11% | 5% | 1% | 0% | 0% | **99,99%** | 9% | 4% | 1% | 0% | 0% |

**Table 6 Credibility for 6M, 1Y, 18M and 2Y Time Horizon**

Furthermore, it is possible to use those weights to construct a Confidence Interval in as much we know the probability distribution they come from and they are probability that can be interpreted as confidence coefficients. In detail we have:

$$\text{Prob}(a \leq f(\theta, X_1, \ldots, X_n) \leq b) \geq Z$$

where $Z$ are the weights shown in Table 6, assuming that weights higher than 100% are weights equals 100%, and $\theta$ are the pvalues shown in Table 3. The benefit of the Uniform distribution $(0,1)$ is that cumulative density function equals the probability density function, therefore the Confidence Interval can be easily obtained:

$$IC = \text{pvalues} * (\pm Z)$$

Table 7 shows the pvalues adjusted for the lower Confidence Interval. We use the pvalues in Table 3 fixing the Probability of $P = 90\%$ and $k = 5\%$ of Table 6. We deem reliable to accept/reject the pvalues according to the standard warning level where above 1% (red) are those estimation to accept.

| Z=alfa | 100% | 100% | 50% | 24% | 12% | 7% | 6% | | | | | | | |
|---|---|---|---|---|---|---|---|---|---|---|---|---|---|---|
| 1-alfa | 0% | 0% | 50% | 76% | 88% | 93% | 94% | | | | | | | |
| **Risk Factor** | **Pvalues for Time Horizons** | | | | | | | **Sample of Time Horizons** | | | | | | |
| **Interest rates** | 2w | 1m | 3m | 6m | 1y | 18m | 2y | 2w | 1m | 3m | 6m | 1y | 18m | 2y |
| ZEROEUR | 0,2% | 3,3% | 1,0% | 2,4% | 0,6% | 0,1% | 0,1% | 137 | 136 | 45 | 22 | 11 | 6 | 5 |
| ZEROEUR-06M | 11,0% | 0,5% | 4,7% | 0,6% | 1,2% | 0,0% | 0,0% | 137 | 136 | 45 | 22 | 11 | 6 | 5 |
| ZEROEUR-12M | 3,4% | 0,3% | 1,3% | 0,4% | 0,8% | 0,1% | 0,1% | 137 | 136 | 45 | 22 | 11 | 6 | 5 |
| ZEROEUR-24M | 6,9% | 0,5% | 1,0% | 0,4% | 1,2% | 0,6% | 0,7% | 137 | 136 | 45 | 22 | 11 | 6 | 5 |
| ZEROEUR-03Y | 4,4% | 0,4% | 0,8% | 0,3% | 1,1% | 0,8% | 0,6% | 137 | 136 | 45 | 22 | 11 | 6 | 5 |
| ZEROEUR-05Y | 4,3% | 0,3% | 0,6% | 0,6% | 1,1% | 0,8% | 0,7% | 137 | 136 | 45 | 22 | 11 | 6 | 5 |
| ZEROEUR-10Y | 4,2% | 0,7% | 0,6% | 0,8% | 1,0% | 0,5% | 0,5% | 137 | 136 | 45 | 22 | 11 | 6 | 5 |
| ZEROEUR-15Y | 6,0% | 2,8% | 2,2% | 2,7% | 2,0% | 0,6% | 0,6% | 137 | 136 | 45 | 22 | 11 | 6 | 5 |
| ZEROEUR-20Y | 10,4% | 4,4% | 3,7% | 4,0% | 3,4% | 0,7% | 0,8% | 137 | 136 | 45 | 22 | 11 | 6 | 5 |
| ZEROEUR-30Y | 5,5% | 9,3% | 6,4% | 2,6% | 4,5% | 0,5% | 1,0% | 137 | 136 | 45 | 22 | 11 | 6 | 5 |
| ZEROEUR-50Y | 12,3% | 7,3% | 11,9% | 0,6% | 4,4% | 0,3% | 1,6% | 137 | 136 | 45 | 22 | 11 | 6 | 5 |

**Table 7 AD pvalues adjusted for Credibility weights**

Table 7 is strongly rejecting the modeling forecast. This is mainly due to the very low credibility weights. Longley-Cook were one of the first researchers addressing the theme of credibility and propose the following alternative method

$$Z = \frac{(1+\gamma)n}{n+\gamma N}$$

where n are the observations in the sample (Table 3), and N are the ones determined with the Full Credibility Criterion (Table 5) $\gamma = 30\%$ as suggested by Longley-Cook. Table 8 shows the credibility of EUR interest rate sample using the Longley-Cook weight, where it is shown that with respect to Table 6 the credibility is higher.

| Time Horizon 1-Month (n=136) | | | | | | Time Horizon 3-Months (n=45) | | | | | |
|---|---|---|---|---|---|---|---|---|---|---|---|
| Prob/k | k =30 % | k =20 % | k =10 % | k =5 % | k =1 % | Prob/k | k =30 % | k =20 % | k =10 % | k =5 % | k =1 % |
| 80% | 128% | 126% | 116% | 88% | 10% | 80% | 125% | 119% | 95% | 53% | 3,5% |
| 85% | 128% | 125% | 113% | 81% | 8% | 85% | 123% | 117% | 89% | 46% | 2,8% |
| 90% | 127% | 124% | 108% | 72% | 6% | 90% | 122% | 113% | 81% | 38% | 2,1% |
| 95% | 126% | 121% | 101% | 61% | 4% | 95% | 119% | 107% | 70% | 29% | 1,5% |
| 97,5% | 125% | 119% | 95% | 52% | 3% | 97,5% | 115% | 102% | 62% | 24% | 1,2% |
| 99% | 123% | 116% | 87% | 44% | 3% | 99% | 111% | 95% | 53% | 19% | 0,9% |
| 99,99% | 116% | 102% | 61% | 24% | 1% | 99,99% | 95% | 71% | 30% | 9% | 0,4% |

| Time Horizon 6-Months (n=22) | | | | | | Time Horizon 1-Year (n=11) | | | | | |
|---|---|---|---|---|---|---|---|---|---|---|---|
| Prob/k | k =30 % | k =20 % | k =10 % | k =5 % | k =1 % | Prob/k | k =30 % | k =20 % | k =10 % | k =5 % | k =1 % |
| 80% | 120% | 109% | 74% | 33% | 2% | 80% | 112% | 94% | 52% | 19% | 0,9% |
| 85% | 117% | 106% | 67% | 27% | 1% | 85% | 107% | 89% | 45% | 15% | 0,7% |
| 90% | 114% | 99% | 58% | 22% | 1% | 90% | 102% | 80% | 38% | 12% | 0,5% |
| 95% | 109% | 91% | 47% | 16% | 1% | 95% | 94% | 69% | 29% | 9% | 0,4% |
| 97,5% | 103% | 83% | 40% | 13% | 1% | 97,5% | 86% | 61% | 23% | 7% | 0,3% |
| 99% | 97% | 74% | 32% | 10% | 0% | 99% | 77% | 52% | 18% | 5% | 0,2% |
| 99,99% | 74% | 48% | 16% | 5% | 0% | 99,99% | 51% | 29% | 9% | 2% | 0,1% |

| Time Horizon 2-Years (n=5) | | | | | | Time Horizon 2-Years (n=5) | | | | | |
|---|---|---|---|---|---|---|---|---|---|---|---|
| Prob/k | k =30 % | k =20 % | k =10 % | k =5 % | k =1 % | Prob/k | k =30 % | k =20 % | k =10 % | k =5 % | k =1 % |
| 80% | 100% | 76% | 35% | 11% | 0% | 80% | 96% | 71% | 30% | 9% | 0% |
| 85% | 93% | 70% | 29% | 9% | 0% | 85% | 88% | 64% | 25% | 7% | 0% |
| 90% | 87% | 60% | 24% | 7% | 0% | 90% | 81% | 55% | 20% | 6% | 0% |
| 95% | 76% | 50% | 18% | 5% | 0% | 95% | 71% | 45% | 15% | 4% | 0% |
| 97,5% | 67% | 42% | 14% | 4% | 0% | 97,5% | 61% | 37% | 12% | 3% | 0% |
| 99% | 58% | 35% | 11% | 3% | 0% | 99% | 52% | 30% | 9% | 2% | 0% |
| 99,99% | 34% | 18% | 5% | 1% | 0% | 99,99% | 30% | 15% | 4% | 1% | 0% |

**Table 8 Credibility using Longley-Cook weights**

As a consequence the AD pvalues adjusted for the lower Confidence Interval using the new credibility shows a better results (Table 9).

| Z=alfa | 100% | 100% | 81% | 58% | 38% | 24% | 20% | | | | | | | |
|---|---|---|---|---|---|---|---|---|---|---|---|---|---|---|
| 1-alfa | 0% | 0% | 19% | 42% | 62% | 76% | 80% | | | | | | | |
| Risk Factor | Pvalues for Time Horizons | | | | | | | Sample of Time Horizons | | | | | | |
| Interest rates | 2w | 1m | 3m | 6m | 1y | 18m | 2y | 2w | 1m | 3m | 6m | 1y | 18m | 2y |
| ZEROEUR | 0,2% | 3,3% | 1,7% | 5,8% | 1,7% | 0,4% | 0,5% | 137 | 136 | 45 | 22 | 11 | 6 | 5 |
| ZEROEUR-06M | 11,0% | 0,5% | 7,6% | 1,4% | 3,7% | 0,1% | 0,2% | 137 | 136 | 45 | 22 | 11 | 6 | 5 |
| ZEROEUR-12M | 3,4% | 0,3% | 2,2% | 1,0% | 2,4% | 0,2% | 0,2% | 137 | 136 | 45 | 22 | 11 | 6 | 5 |
| ZEROEUR-24M | 6,9% | 0,5% | 1,6% | 0,9% | 3,7% | 2,2% | 2,5% | 137 | 136 | 45 | 22 | 11 | 6 | 5 |
| ZEROEUR-03Y | 4,4% | 0,4% | 1,4% | 0,8% | 3,3% | 2,8% | 2,3% | 137 | 136 | 45 | 22 | 11 | 6 | 5 |
| ZEROEUR-05Y | 4,3% | 0,3% | 1,1% | 1,5% | 3,4% | 2,9% | 2,5% | 137 | 136 | 45 | 22 | 11 | 6 | 5 |
| ZEROEUR-10Y | 4,2% | 0,7% | 1,1% | 1,8% | 3,2% | 1,7% | 1,7% | 137 | 136 | 45 | 22 | 11 | 6 | 5 |
| ZEROEUR-15Y | 6,0% | 2,8% | 3,6% | 6,4% | 6,2% | 2,2% | 2,1% | 137 | 136 | 45 | 22 | 11 | 6 | 5 |
| ZEROEUR-20Y | 10,4% | 4,4% | 6,0% | 9,4% | 10,4% | 2,6% | 3,0% | 137 | 136 | 45 | 22 | 11 | 6 | 5 |
| ZEROEUR-30Y | 5,5% | 9,3% | 10,5% | 6,2% | 13,9% | 1,8% | 3,8% | 137 | 136 | 45 | 22 | 11 | 6 | 5 |
| ZEROEUR-50Y | 12,3% | 7,3% | 19,4% | 1,4% | 13,5% | 1,2% | 5,8% | 137 | 136 | 45 | 22 | 11 | 6 | 5 |

**Table 9 AD pvalues adjusted for Credibility using Longley-Cook weights**